\documentclass[conference]{IEEEtran}
\IEEEoverridecommandlockouts

\usepackage{cite}
\usepackage{amsmath,amssymb,amsfonts}
\usepackage{algorithmic}
\usepackage{graphicx}
\usepackage{textcomp}

\usepackage{lineno,hyperref}
\modulolinenumbers[5]

\begin{document}


\title{Violence originated from Facebook: A case study in Bangladesh
}

\author{\IEEEauthorblockN{Matiur Rahman Minar}
\IEEEauthorblockA{\textit{Department of Computer Science and Engineering} \\
\textit{Chittagong University of Engineering and Technology}\\
Chittagong-4349, Bangladesh \\
minar09.bd@gmail.com}
\and
\IEEEauthorblockN{Jibon Naher}
\IEEEauthorblockA{\textit{Department of Computer Science and Engineering} \\
\textit{Chittagong University of Engineering and Technology}\\
Chittagong-4349, Bangladesh \\
jibon.naher09@gmail.com}
}

\maketitle

\begin{abstract}
Facebook as in social network is a great innovation of modern times. Among all social networking sites, Facebook is the most popular social network all over the world. Bangladesh is no exception. People use Facebook for various reasons e.g. social networking and communication, online shopping and business, knowledge and experience sharing etc. However, some recent incidents in Bangladesh, originated from or based on Facebook activities, led to arson and violence. Social network i.e. Facebook was used in these incidents mostly as a tool to trigger hatred and violence. This case study discusses these technology related incidents and recommends possible future measurements to prevent such violence.
\end{abstract}




\section{Introduction}
\label{S:1}

Almost every single person using internet nowadays uses social networking sites (SNS). Facebook is the most used SNS of them all. Facebook was originally developed as a communicating site for college students \cite{Facebook:2017wk}. Today, people use Facebook for lots of reasons, e.g. communicating, online shopping, marketing, studying, sharing etc. Currently, s More than two billion people actively use Facebook worldwide \cite{Fbinfo:2017wk}In Bangladesh, the number of people use Facebook is 26 million \cite{Bdfb:2017bd}. 

Facebook is the most popular social network in Bangladesh \cite{Bhabit:2017bd}. Almost all internet users in Bangladesh use Facebook \cite{Binternet:2017bd} regardless of age, gender, locality or ethnic identity \cite{BinMorshed:2017:IGD:2998181.2998237}. Also Dhaka is ranked second in top cities for total number of active Facebook users \cite{Bdfb:2017bd}. Facebook is now a integral part of daily activities. Bangladeshi people mainly use Facebook for social Communication, knowledge and experience sharing, and business purposes \cite{Bhabit:2017bd}. 

\begin{figure}[hbtp]
\label{rangpur}
\includegraphics[scale=0.17]{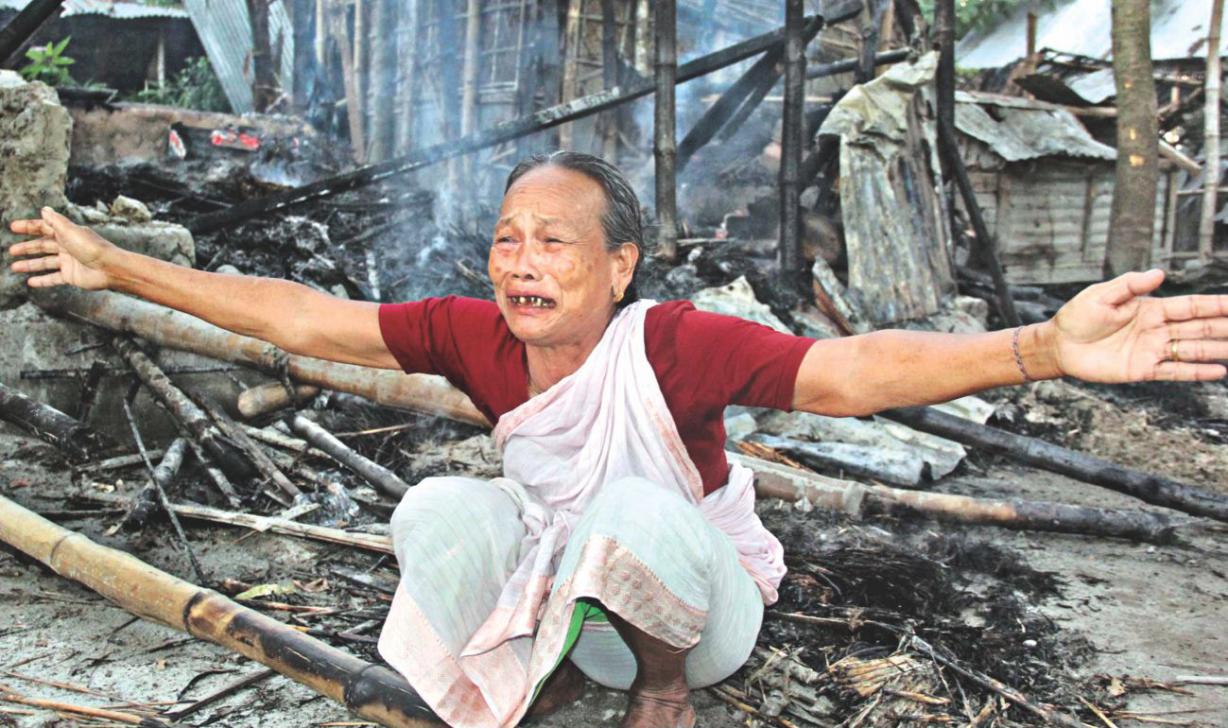}
\includegraphics[scale=0.2]{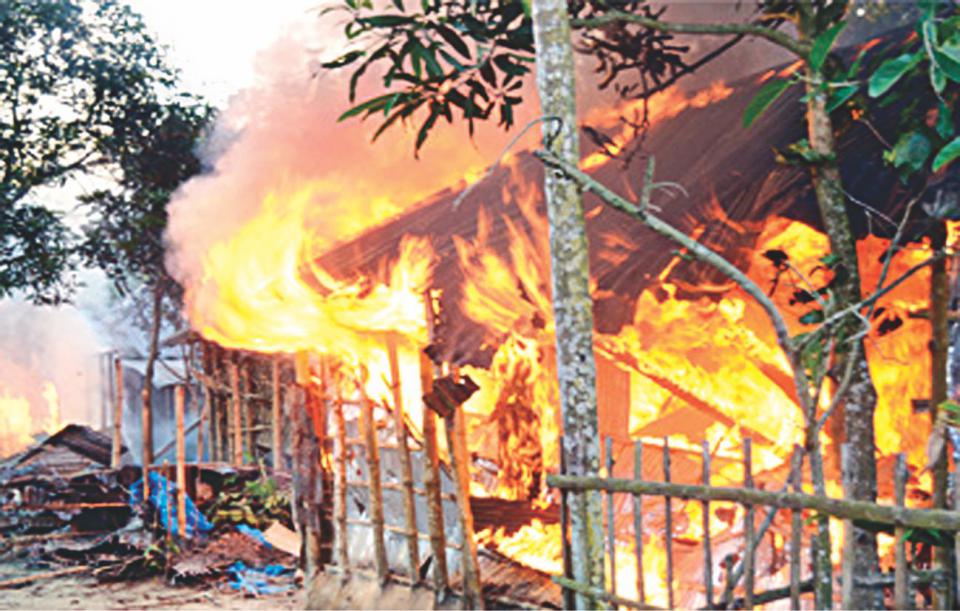}\\
\includegraphics[scale=0.36]{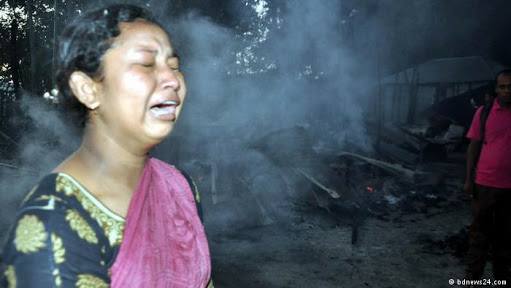}
\includegraphics[scale=0.28]{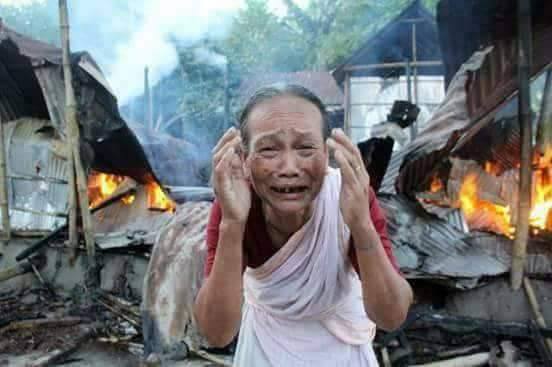} 
\caption{Rangpur 2017}
\end{figure}

Since all sorts of people in Bangladesh are using Facebook for social networking and other purposes, its undeniable that opposite kind of uses of social networks are bound to happen. Starting from social scams, business frauds and others, worst kind of extremism and violence are being triggered from Facebook activities. Some recent incidents made it worse than before. First major incident was Ramu 2012 (\autoref{ramu}) and the latest one is Rangpur 2017 (\autoref{rangpur}). In this study, we are going to briefly discuss about the latest incidents. These are results of arson and violence, intrigued from social network i.e. Facebook activities.

We are going to discuss the incidents in section \ref{S:2}. In section \ref{S:3}, we will analyze the cases, discuss the similarities and the links with social network i.e. Facebook. Section \ref{S:4} is about some possible recommendations about how we can prevent these kind of violence from technology perspectives.

\section{Case overview}
\label{S:2}

In recent 5-6 years, quite a number of incidents happened, where Facebook was the trigger to initiate the violent situations. We are going to discuss these cases in descending manner, starting from the latest. Relevant pictures of these violence are added, which are collected from referenced news articles.

\subsection{Rangpur, 2017} 
In November 10, 2017, a clash broke out in Thakurpara, Rangpur \cite{DailyStar:2017clash}. The clash is said to be triggered by a Facebook post \cite{DailyStar:2017mayhem}. This resulted 1 killed, 20 hurt including 7 policemen \cite{DailyStar:2017clash} \cite{DailyStar:2017mayhem} \cite{BdNews:2017saved}. At least 30 Hindu houses were burned and vandalized in Horkoli Thakurpara village of Rangpur \cite{DailyStar:2017mayhem} \cite{BdNews:2017saved} as shown in figure 1.

This incident was triggered by a Facebook post. In November 5, 2017, Titu Chandra Roy from Rangpur, Bangladesh who was recently living in Narayanganj, Bangladesh shared a Facebook status said to be “Defaming religion” \cite{DailyStar:2017mayhem}. Similar to Ramu violence in 2012 \cite{Wikipedia:2012ramu}, a group of arsonists put fire to the property of Hindu minority at Horkoli Thakurpara village of Rangpur \cite{DailyStar:2017mayhem}. The controversial Facebook post being ‘defamation of Prophet Muhammad’, triggered the communal attack on the Hindu houses by irrate mob in Thakurpara village of Rangpur \cite{HinduExistence:2017rangpur}. The Facebook account resembling Titu Roy is named as MD Titu. The account was opened just two months before of the incident, in September 2017, although it had managed to make 288 Facebook friends by this time \cite{HinduExistence:2017rangpur}.

Sirajul Islam, an Imam of a mosque in a nearby village, and Alamgir Hossain, a trader, filed a case against one Titu Chandra Roy with Gangachara Police Station, accusing him of making a post on Facebook that hurt religious sentiments of Muslims \cite{DailyStar:2017framed} \cite{DailyStar:2017accused}. Then the influential locals spread hatred which resulted the violence and clash with police \cite{DailyStar:2017framed}. A mobile phone number is found which was used for opening the Facebook ID around two months ago. The phone number was found unused and police could not verify who the number was registered to. The owner of the Facebook account generally shared posts or uploaded screenshots of others' posts \cite{DailyStar:2017framed} \cite{DailyStar:2017accused}. Tito Chandra Roy, an accused in a case filed for allegedly making a post on Facebook that sparked mayhem in Thakurpara village of Rangpur, was arrested in Jaldhaka upazila of Nilphamari district in November 14, 2017 \cite{DailyStar:2017accused} \cite{BdNews:2017arrested}. 
Ziton Bala, mother of Tito Chandra Roy, said \cite{DailyStar:2017framed} \cite{DailyStar:2017accused}, 
\begin{quote}
\textit{My son's name is Tito; people are telling me his name is Titu and it's not true. }
\end{quote}
The home minister said \cite{BdNews:2017arrested}, 
\begin{quote}
\textit{We have heard that Titu is illiterate }\cite{HinduExistence:2017rangpur}.
\end{quote}
Rangpur Superintendent of Police Mizanur Rahman said  \cite{BdNews:2017arrested},
\begin{quote}
\textit{We did not find any status insulting religion on his Facebook. False information was spread to stir attacks, arson and looting against the Hindu community.}
\end{quote}

\subsection{Brahmanbaria, 2016}
In October, 2016, violent mob carried out a synchronized attack on the Hindus in Brahmanbaria's Nasirnagar upazila, destroying and setting fire on more than 150 homes and at least 15 temples and looting valuables \cite{HinduExistence:2016bbaria} over an alleged Facebook post insulting Islam \cite{BdNews:2016withdrawn} \cite{DailyStar:2016police}, as shown in figure 2. At least 20 people including several temple devotees were left wounded in the attack \cite{BdNews:2016withdrawn}. Later, twice more, attacks were carried out on Hindus – setting their houses on fire \cite{DailyStar:2016police} \cite{HinduExistence:2016bbaria} \cite{DailyStar:2016planned}.

The violence was triggered by a Facebook post purportedly from the account named 'Rasraj Das', son of Jagannath Das at Haripur Union’s Harinberh village for "hurting religious sentiments of Muslims", as the locals said \cite{BdNews:2016withdrawn}. Police later arrested Rasraj for ‘denigrating Islam’ through his post on the social media. A court then ordered him into prison \cite{BdNews:2016withdrawn} \cite{DailyStar:2016suspect} \cite{Breakingnews:2017two} \cite{DailyStar:2016planned}.

It was later exposed that Awami League leader Faruk Mia, the District Union President of Nasirnagar had some problem with the local Fishermen Union leader Rasaraj Das \cite{DailyStar:2016planned}. Hence, Faruk opened an Facebook account in the name of Rasaraj Das \cite{HinduExistence:2016bbaria}.Then, Faruk posted a picture of Kaba juxtaposed with Hindu deity Lord Shiva with the help of his brother Kaptan Mia in the timeline of Rasaraj \cite{HinduExistence:2016bbaria}.

\begin{figure}[hbtp]
\includegraphics[scale=0.33]{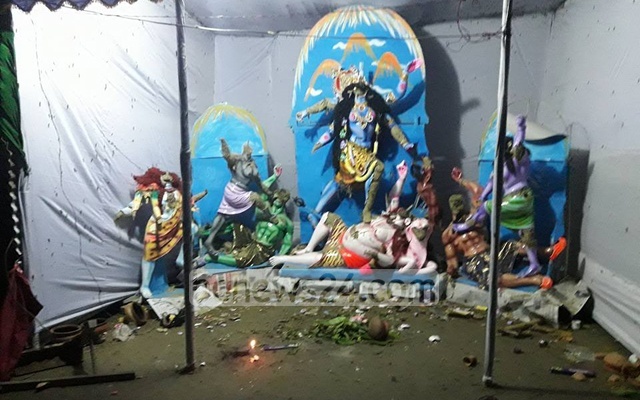}
\includegraphics[scale=0.35]{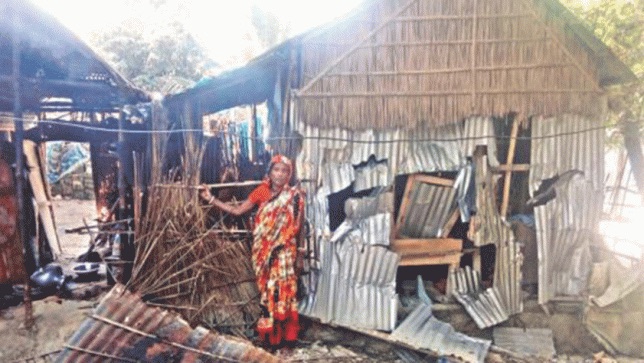}
\caption{Brahmanbaria 2016}
\end{figure}

\subsection{Comilla, 2014}
In April, 2014, fake news of defaming religion on Facebook triggered violent attacks over Hindu minority in Homna Upazilla of Comilla, Bangladesh \cite{HinduExistence:2014comilla} as shown in figure 3. At least 28 houses of Hindus were ransacked at Bakhsitarampur village in Homna upazila of Comilla in an attack of a mob of nearly 3,000 prompted by rumours that Prophet Muhammad (pbuh) had been defamed in Facebook posts by some Hindus \cite{DailyStar:2014rumour} \cite{BdChronicle:2014attack} \cite{HinduSamhati:2014smashed} \cite{HinduExistence:2014comilla}. Homna police station OC Aslam Shikder said that hundreds of people from the Panchkipta village attacked temples and homes of at least 28 Hindu families alleging that two Hindu youths posted defamatory comments on Facebook about the Prophet \cite{BdChronicle:2014attack}. Police detained two persons Utshab Das and Srinibas Das as well as nine others over the rumours of Facebook post  \cite{BdChronicle:2014attack}. Homna police station OC Aslam Shikder said,
\begin{quote}
\textit{During the interrogation they have denied posting any such remark.} \cite{BdChronicle:2014attack}
\end{quote}

Villagers said a call was made from the loudspeakers of a Madrasa at Rampur village, near Baghsitarampur, to launch the assault on the Hindus. Before the attacks, leaflets were distributed for last several days in the madrasas claiming that two Hindu youths had slandered the prophet in a Facebook post on April 27, 2014. \cite{BdNews:2014madrasa}

\begin{figure}[hbtp]
\centering
\includegraphics[scale=0.22]{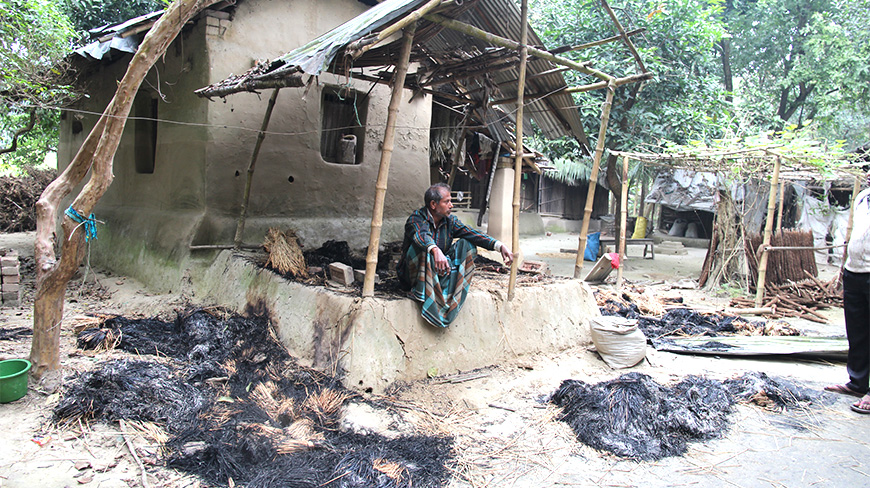}
\includegraphics[scale=0.3]{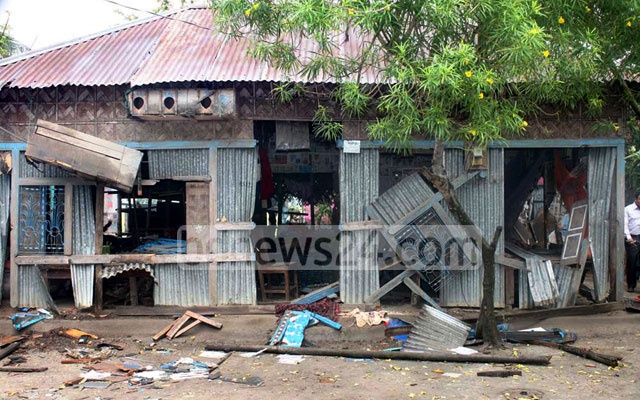}
\caption{Comilla 2014}
\end{figure}

\subsection{Pabna, 2013}
In November 3, 2013, A mob went on a rampage in a Hindu-dominated neighbourhood in Bonogram of Pabna, Bangladesh following reports that a boy from the minority community had committed blasphemy. As a result, more than 25 houses belonging to Hindus vandalised as shown in figure 4, several idols in temples damaged and about 150 families forced to flee the area  \cite{Persecution:2013mob} \cite{HinduExistence:2013pabna} \cite{DailyStar:2013attacked}.

A group of people were distributing photocopies of what they said was a “Facebook page”. They claimed one Rajib Saha had maligned Prophet Mohammad (pbuh) in the page \cite{DailyStar:2013attacked} \cite{HinduHumanRights:2013false}. An eyewitness said,
\begin{quote}
\textit{None was given the chance to ask whether or not it was a faked Facebook posting.}\cite{DailyStar:2013attacked}
\end{quote}
Rajib, son of Babul Saha, a shop owner in the bazaar, is a class-X student of Bonogram Miapur High School \cite{DailyStar:2013attacked}. Babul Saha said,
\begin{quote}
\textit{He was preparing for SSC examination. He can't do anything like what the people here are alleging.}\cite{DailyStar:2013attacked}
\end{quote}

The name of the Facebook page was written in Bangla and denigrates the prophet. It was opened on September 14, 2013. It did contain hateful posts, majority of which was issued by the administrator. Some people gave negative reactions to these posts, asking others to refrain from liking this page \cite{DailyStar:2013attacked} \cite{HinduHumanRights:2013false}. It was later found that the Facebook page whose photocopies were used to incite the attacks on the Bonogram Hindu community has no links with Rajib \cite{DailyStar:2013attacked} \cite{HinduExistence:2013pabna}.

\begin{figure}[hbtp]
\centering
\includegraphics[scale=0.28]{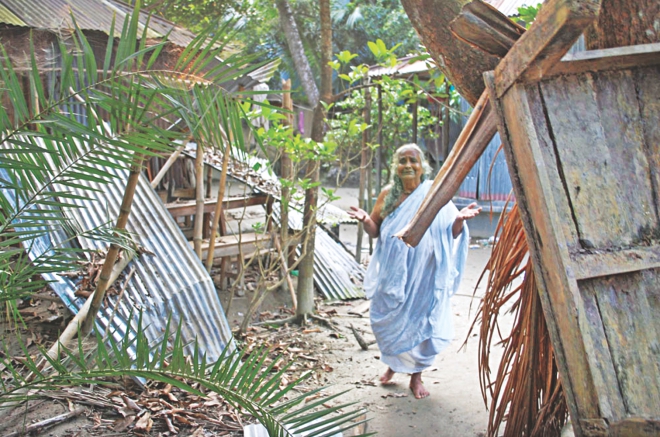}
\includegraphics[scale=0.28]{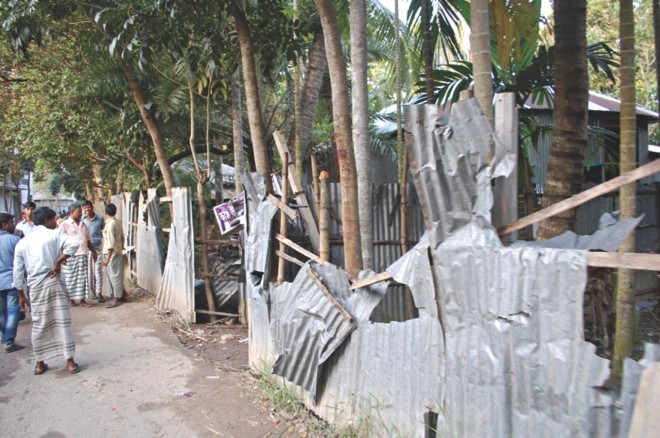} 
\caption{Pabna 2013}
\end{figure}

\subsection{Ramu, 2012}
Ramu is an Upazilla of Cox's Bazar District in Bangladesh. It is located in the  south-east area of the country, under the division of Chittagong. According to tradition, Ramu, Cox's Bazar got its name from the Ramu dynasty of the Arakan. It came under the Mughals when Chittagong was captured by them and during that time a thirteen-foot-high Buddha statue was found. 
Buddhist temples in Ramu, Cox’s Bazar is of great importance towards Buddhists as well as tourists. \cite{Wikipedia:2017upazilla} \cite{Wikipedia:2017ramu}

In September 2012, Buddhist temples in Ramu, Cox’s Bazar was burnt to the ground. It was originated from a local Buddhist, Uttam Kumar Barua being tagged in a Facebook image of Quran. An unknown/fake Facebook user, using a pseudonym, posted burning-Quran image on Uttam Kumar Barua’s Facebook wall. Reacted by the post, a group of arsonists put fire at the Buddhist temple. Fanatics attacked the Buddhist community in Cox's Bazar's Ramu, claiming that a Buddhist youth “insulted Islam” on social media. \cite{Wikipedia:2012ramu} \cite{DailyStar:2012linked} \cite{BBC:2012facebook} \cite{BdNews:2012ukhiya} \cite{Archive:2012mob} \cite{Archive:2012burned} \cite{DailyStar:2012soul}

The mobs destroyed 12 Buddhist temples and monasteries and 50 houses as shown in figure 5. The violence started in reaction to a tagging of an image depicting the desecration of a Quran on the timeline of a fake Facebook account under a Buddhist male name. The actual posting of the photo was not done by the Buddhist who was falsely slandered.  The Buddhist was innocent of the accusation. The violence later spread to Ukhiya Upazila in Cox's Bazar District and Patiya Upazila in Chittagong District where Buddhist monasteries, Sikh Gurudwaras and Hindu temples were targeted for attacks. \cite{Wikipedia:2012ramu} \cite{DailyStar:2012linked} \cite{BBC:2012facebook} \cite{BdNews:2012ukhiya} \cite{Archive:2012mob} \cite{Archive:2012burned} \cite{HinduExistence:2012ramu}

The man who sparked the riots, who has gone into hiding, told local media he did not post the picture, insisting someone else had "tagged" his account with the image on the social network \cite{Archive:2012mob}. Later, it was found that, the Facebook page with an anti-Islam picture that provoked the September 29, 2012 rampage against the Buddhist community in Ramu was photoshopped \cite{DailyStar:2012devil}. Somebody or a group had taken a screenshot of Uttam Kumar Barua's facebook profile page, cut out the address of anti-Islam website “Insult allah” and pasted it on the address bar visible in the image. Once the fabrication was done, it looked like the website has shared the anti-Islam image with Uttam and 26 others. \cite{DailyStar:2012devil} \cite{HinduExistence:2012ramu}

\begin{figure}[hbtp]
\label{ramu}
\centering
\includegraphics[scale=0.5]{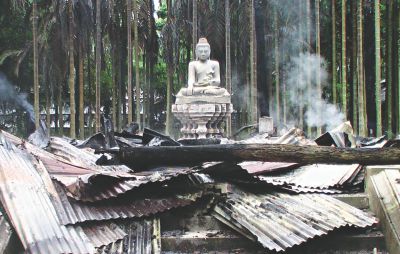} 
\includegraphics[scale=0.26]{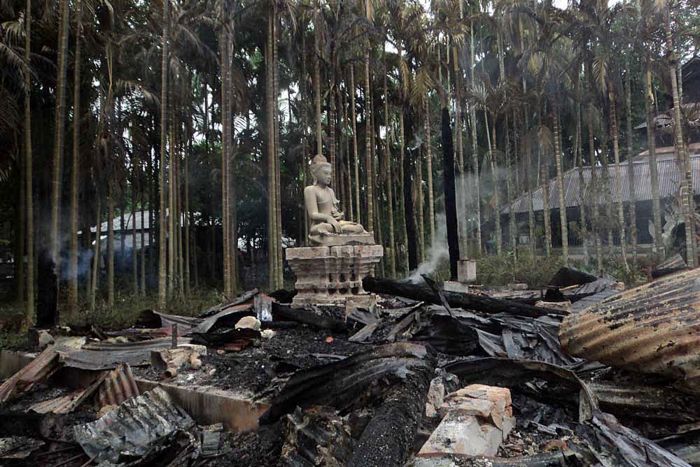}\\
\includegraphics[scale=0.28]{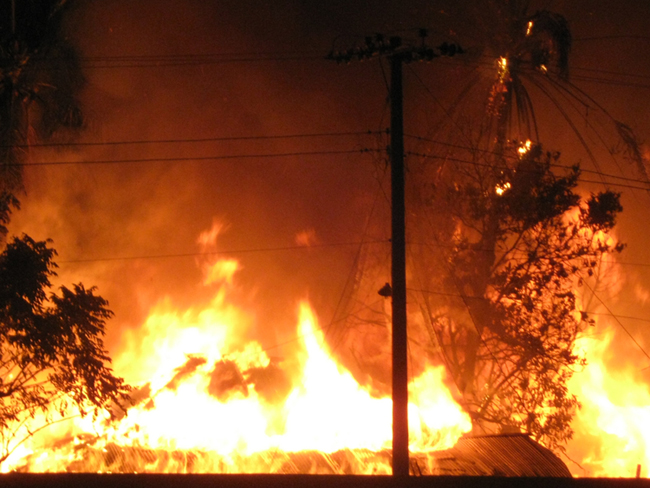} 
\includegraphics[scale=0.5]{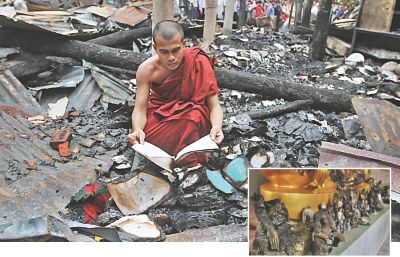}
\caption{Ramu 2012}
\end{figure}

\section{Analysis and Discussions}
\label{S:3}
Bangladesh saw dozen of virulent attacks on minorities with spreading rumors or posting objectionable elements on social media against Islam.
In these cases, perpetrators took the advantage of popular social media platform Facebook to generate fundamental sentiments to attack minorities in Bangladesh \cite{HinduExistence:2016bbaria} \cite{HinduExistence:2017rangpur}.

From these cases, one thing is common. All started from Facebook posts, which triggered arson and violence. Also, the arsonists and attackers seem to be greatly affected by social network i.e. Facebook posts/activity. This indicates the availability and expansion of social networks in a third-world and developing country like Bangladesh. This also indicates the ignorance and unwillingness of local/related people towards verification of authenticity of such defaming posts/activities. Facebook activities are seem to be taken into granted here without any context or authenticity. Also it seems some are taking advantages of this ignorance, trying deliberately to drive people towards hatred and extremism, while gaining different purposes.

\section{Possible future prevention}
\label{S:4}
Many research and development works are being done using technology oriented solutions for Bangladeshi social problems \cite{Ahmed:2015:RMI:2702123.2702573}. For example, Protibadi \cite{Ahmed:2014:PPF:2611205.2557376} is a notable one. Also there are works like remote health monitoring \cite{Minar:2017vital} in the context of Bangladesh. \cite{MBGDMQN2013} studied some trust issues of social networks and proposed ways to solve them.

Considering this study, social networks need to be stricter on moderating users’ identity and activities. It seems that some people are deliberately using Facebook as a medium of triggering violence. There should be multiple levels of identity verification of the users and their purposes in social networks. They may take suitable steps to identify fake/inactive users and close their accounts as well.

To prevent these kind of violence beforehand, social networks and defensive forces can use detection system over Facebook and other social networks. Detecting suspicious users and activities from social networks could be useful in this regard. For example, Facebook recently launched Artificial Intelligence based program to predict and prevent suicides \cite{Wired:2017suicide}. Similar programs can be implemented to predict and prevent violence and extremism.





\bibliographystyle{IEEEtran}
\bibliography{violence-from-facebook.bib}







\end{document}